\documentclass[aps,prd,preprint,nofootinbib]{revtex4-2}
\usepackage{amsmath,amsfonts,amssymb,amsbsy}
\usepackage{mathrsfs,latexsym}
\usepackage{graphicx}
\usepackage{subcaption}
\usepackage{xcolor}
\usepackage{float}
\usepackage{dcolumn}% Align table columns on decimal point
\usepackage[bookmarks,
bookmarksopen = true,
bookmarksnumbered = true,
linktocpage,
colorlinks = true,
linkcolor = blue,
urlcolor  = blue,
citecolor = blue,
anchorcolor = green,
hyperindex = true,
hyperfigures]{hyperref}
\usepackage{array}
\usepackage{booktabs}
\usepackage{multirow}
\usepackage{placeins}
\usepackage{blindtext}
\usepackage{appendix}
\usepackage{fontawesome} % requires XeTeX or LuaTeX
\usepackage{bm}
\usepackage{times}
\usepackage{float}
\usepackage{wrapfig}
\usepackage{physics}
\usepackage[utf8]{inputenc} % Required for inputting international characters
\usepackage[T1]{fontenc} % Output font encoding for international characters
\usepackage{amsmath,amssymb}

\newcommand{\ben}{\begin{enumerate}}
\newcommand{\een}{\end{enumerate}}
\newcommand{\bit}{\begin{itemize}}
\newcommand{\eit}{\end{itemize}}
\newcommand{\beq}{\begin{equation}}
\newcommand{\eeq}{\end{equation}}
\newcommand{\bsa}{\begin{subequations}\begin{eqnarray}}
\newcommand{\esa}{\end{eqnarray}\end{subequations}}
\newcommand{\bea}{\begin{eqnarray}}
\newcommand{\eea}{\end{eqnarray}}
\newcommand{\bean}{\begin{eqnarray*}}
\newcommand{\ean}{\end{eqnarray*}}

\newcommand{\sbullet}[1][.5]{\mathbin{\vcenter{\hbox{\scalebox{#1}{$\bullet$}}}}}

\usepackage{xcolor, comment}

\usepackage[justification=raggedright,singlelinecheck=false]{caption}

\begin{document}
\begin{flushright}
    \texttt{DESY-25-184}\\
    \texttt{CERN-TH-2025-249}
\end{flushright}

\title{Investigating a two-level algorithm for fermionic observables}%

\author{Lorenzo Barca, Stefan Schaefer}
\affiliation{John von Neumann-Institut für Computing NIC,
Deutsches Elektronen-Synchrotron DESY, Platanenallee 6, 15738 Zeuthen, Germany}

\author{Jacob Finkenrath}
\affiliation{Department of Physics, Bergische Universität Wuppertal, Gaußstr. 20, 42119 Wuppertal, Germany}
\affiliation{CERN, Esplanade des Particules 1, 1211 Geneva 23, Switzerland}
\collaboration{FOR5269 - "Future methods for studying confined gluons in QCD"}

\begin{abstract}
We investigate the combination of a two-level sampling algorithm with distillation techniques 
to compute disconnected fermionic correlation functions. The method relies on a factorization 
of the quark propagator into domain-local contributions that depend only on the gauge fields 
within overlapping temporal regions, enabling independent submeasurements of each term through 
a two-level sampling strategy. The two-level estimators exhibit the expected $1/N_1^2$ scaling 
of the variance, up to exponential boundary effects, and achieve an exponential reduction of 
statistical errors at nearly the same computational cost as standard sampling. The method is 
tested on pure gauge ensembles, providing a controlled benchmark for its forthcoming application 
to dynamical QCD studies of glueball and isosinglet meson correlation functions.
\end{abstract}

\maketitle

\section{\label{sec:intro}Introduction}
One ubiquitous challenge in lattice QCD simulations is the exponential degradation of the signal-to-noise ratio in correlation functions \cite{Parisi:1983ae, Lepage:1989hd}.
A solution to this problem would eliminate one of the most severe limitations currently hindering the precision and scope of lattice QCD calculations.
For hadron structure observables, such as nucleon form factors \cite{Barca:2025det} and semileptonic decay amplitudes of heavy-light mesons \cite{FermilabLattice:2022gku}, the exponential degradation of the signal-to-noise ratio restricts the accessible source-sink separations, limiting control over excited-state contamination. 
At the same time, high-precision determinations of correlation functions at long Euclidean distances are crucial for phenomenologically important observables, such as the lattice QCD determination of the hadronic contributions to the muon $g-2$~\cite{FermilabLatticeHPQCD:2024ppc}.
The signal-to-noise problem is particularly severe in the study of glueballs, which suffer from notoriously poor signal-to-noise, see Refs.~\cite{Vadacchino:2023vnc, Morningstar:2024vjk} for recent reviews, and are the central focus of this work.

In particular, the lattice computation of glueball observables is exacerbated because of disconnected contributions: while the signal decays exponentially with the distance between the operators, the statistical noise remains constant with standard sampling techniques.
Nonetheless, lattice simulations have provided robust predictions for the glueball spectrum in pure SU(N) Yang-Mills theories \cite{Athenodorou:2020ani, Lucini:2010nv, Chen:2005mg, Lucini:2004my, Morningstar:1999rf, Bali:1993fb}, as well as recent first investigations of their internal structure \cite{Abbott:2025irb}, offering valuable insights into the mass hierarchy, quantum numbers and properties of these states in the absence of quarks.
However, the identification of glueballs in nature remains elusive, as glueball states are expected to mix with conventional mesons and appear in decay channels involving multi-particle final states \cite{Petrov:2022ipv, Klempt:2021wpg, Crede:2008vw}.
Experimental searches for glueballs are still ongoing, particularly in the radiative decays of charmonium states such as the $J/\psi$. 
Notably, recent results include the first identification of a glueball candidate with pseudoscalar quantum numbers \cite{Liu:2024anq}. 

Consequently, realistic and unambiguous determinations of glueball properties in full lattice QCD require correlation functions built from a comprehensive set of interpolating operators, including pure gluonic, mesonic, hybrid, and multi-hadron interpolators. This broad operator basis is essential to capture relevant decay channels, mixing patterns, and coupled-channel dynamics. However, such studies remain extremely challenging: the factorial growth of Wick contractions, the need for all-to-all quark propagators, and the poor signal-to-noise ratio of disconnected contributions at early time separations significantly limit the achievable precision.

Due to these challenges, robust investigations of glueballs in dynamical QCD remain very scarce. The few existing dynamical studies, such as Refs.~\cite{Athenodorou:2023ntf, Gui:2019dtm, Sun:2017ipk, Gregory:2012hu, Richards:2010ck}, rely on relatively small variational bases, in some cases consisting only of pure-gluonic operators, which limit their ability to fully resolve the physics underlying glueball–meson mixing.
Recent works have expanded the variational basis by exploiting distillation \cite{Peardon:2009gh, Morningstar:2011ka}, enabling the inclusion of multi-hadron, fermionic, and pure-gluonic interpolators \cite{Urrea-Nino:2025afu, Jiang:2022ffl, Brett:2019tzr}.
Nonetheless, their precision remains limited by the poor signal-to-noise ratio of disconnected contributions at early times or by the still restricted number of operators in the basis.

Alternative sampling methods, like the multilevel algorithm \cite{Parisi:1983hm, Luscher:2001up, Meyer:2002cd, Meyer:2003hy}, can be used to reduce the error in a more efficient way than standard sampling.
In \cite{Barca:2023arw, Barca:2024fpc}, we demonstrate that a two-level sampling can reduce more efficiently the statistical error of pure gauge glueball two-point functions at large distances.
A key idea behind the two-level algorithm is to exploit locality by splitting the temporal extent of the lattice into active regions separated by frozen regions, defined in \cite{Barca:2024fpc}, and performing $N_1$ independent submeasurements in each active region. When the source and sink operators lie in different active regions, the resulting factorisation leads to a variance scaling ideally as $1/N_1^2$, up to corrections which saturate the two-level error reduction at large $N_1$ and which depend exponentially on the distance of the operators from the frozen regions.

However, the application of multilevel sampling with fermions in full QCD is not straightforward.
After integrating out the fermionic degrees of freedom, the fermion observables are written as traces of products 
of quark propagators times the determinants of the quark propagators, which depend on the gauge fields over the full space-time and therefore hinders the application of multilevel algorithms.
By decomposing the quark propagator in the different regions that comprise the lattice extent, 
it is possible to make the fermionic observables amenable for multilevel integration.
In quenched QCD, an important step forward has been made in \cite{Ce:2016idq, Ce:2016qto} by rewriting the quark propagator as a series of terms with a well defined hierarchical structure. 
This factorisation of the quark propagator enables a two-level integration of the fermionic observables.

In full QCD, the application of multilevel integration requires more advanced techniques due to the presence of the fermion determinant as the fermionic weight.
A factorisation of the fermion determinant via a multiboson approximation \cite{Luscher:1993xx} 
makes the fermion weight amenable for a local integration, as demonstrated in \cite{Ce:2016ajy, Ce:2017ndt, DallaBrida:2020cik}.

As a step towards the computation of the glueball spectrum in full QCD, we combine, for the first time, distillation techniques with a two-level algorithm in quenched QCD. This combined approach is particularly valuable for glueball studies: distillation tackles the factorial growth of Wick contractions arising from the multi-particle operator basis required in comprehensive dynamical analyses, while the two-level sampling mitigates the poor signal-to-noise ratio of disconnected contributions in isosinglet observables. A key question we investigate is whether two-level sampling can improve the signal already at sufficiently short operator separations, where it is already needed but where our pure-gauge study revealed limitations from boundary effects. This work thus provides a benchmark for future applications in full QCD, where a reliable determination of glueball properties requires several correlation functions with high statistical precision, and more broadly paves the way for next-generation studies of hadron correlation functions facing similar signal-to-noise challenges.
\section{Lattice QCD correlation functions and signal-to-noise ratio}\label{sec:sec2}
We are interested in estimating two-point correlation functions of the form
\begin{equation}
\langle \mathrm{O}(\vec{p}, x_0) \bar{\mathrm{O}}(\vec{p}, y_0)\rangle
=
\frac{1}{\mathcal{Z}}\int [dU] [dq] [d\bar{q}] e^{-S[q, \bar{q}, U]}\mathrm{O}(\vec{p}, x_0) \bar{\mathrm{O}}(\vec{p}, y_0)\,,
\end{equation}
where $\mathrm{O}$ denotes interpolating operators constructed to have overlap with specific states (e.g.\ glueballs), 
$\mathcal{Z}$ is the QCD partition function, and $S = S_f + S_g$ the QCD action.
After integrating out the fermionic fields, the correlation function in the quenched approximation becomes
\begin{equation}\label{corr}
\langle \mathrm{O}(\vec{p}, x_0) \bar{\mathrm{O}}(\vec{p}, y_0)\rangle
=
\frac{1}{\mathcal{Z}_g} \int [dU]  e^{-S_g[U]} \langle \mathrm{O}(\vec{p}, x_0) \bar{\mathrm{O}}(\vec{p}, y_0)\rangle_F\,,
\end{equation}
where $\langle \cdots \rangle_F$ denotes the Wick contractions expressed in terms of traces of products of Dirac propagators.
In this work we focus on single-meson operators $O_\Gamma$ and $\pi\pi$ interpolators projected to isospin $I=0$,
i.e., singlets under $\mathrm{SU}(2)$ isospin transformations.
After projecting to zero total momentum, single-meson operators read
\begin{equation}
\label{meson_op}
\mathrm{O}_\Gamma(\vec{0}, x_0)=\sum_{\vec{x}} \bar{q}(x) \Gamma q(x)\,,
\end{equation}
with $x=(\vec{x}, x_0)$, and $\Gamma=I, \gamma_5, \gamma_\mu, \gamma_4\gamma_5, \gamma_i\gamma_j$.
Since in this work we consider only zero momentum, we drop the explicit momentum labels for simplicity.

In particular, the Wick contractions of the two-point functions of $\mathrm{O}=\mathrm{O}_\Gamma$ contain \textit{disconnected quark-line} contributions, which read
\begin{equation}
\label{corrql}
\langle \mathrm{O}(x_0) \bar{\mathrm{O}}(y_0)\rangle_{F, \rm disc}
=
\sum_{\vec{x}, \vec{y}}
\langle \mathrm{O}(x)\rangle_F \langle\bar{\mathrm{O}}(y)\rangle_{F}\,.
\end{equation}
For instance, in the case of single-meson operators, 
the explicit expressions of the Wick contractions contain quark loops, which read
\begin{equation}
\mathcal{O}_\Gamma(x_0)
:=
\sum_{\vec{x}}
\langle \mathrm{O}_\Gamma(x)\rangle_F
=
-
\sum_{\vec{x}}
\mathrm{Tr} \left[D^{-1}(x,x) \Gamma \right]\,.
\end{equation}
For scalar observables such as $\mathcal{O}_I$, the gauge average of the one-point function has a 
non-vanishing vacuum expectation value (VEV), which is constant in time.

The signal is obtained after subtracting this VEV from the two-point correlator:
\begin{equation}
C(x_0-y_0)_{\rm disc} = \langle \mathcal{O}(x_0) \bar{\mathcal{O}}(y_0)\rangle - \langle \mathcal{O}(x_0) \rangle \langle \bar{\mathcal{O}}(y_0)\rangle.
\end{equation}
Correlation functions are estimated in lattice QCD \cite{Wilson:1974sk} 
by generating a finite number $N$ of gauge configurations $U$ according to the QCD probability distribution,
and computing ensemble averages via importance sampling.
The physical signal decays exponentially with the operator separation,
\begin{equation}
C(x_0-y_0)_{\rm disc}  \sim e^{-E|x_0-y_0|}\,,
\end{equation}
while the variance of disconnected quark-line contributions remains constant in time \cite{Lepage:1989hd}.
Consequently, the signal-to-noise ratio deteriorates exponentially \cite{Parisi:1983ae},
\begin{equation}
\abs{C(x_0-y_0)_{\rm disc}} \Bigl/ \sigma_C(x_0-y_0) \sim \sqrt{N} ~e^{-E|x_0-y_0|}\,.
\end{equation}
Therefore, the analysis is restricted to the distances where the signal is still much larger than the noise.

In principle, statistical precision can be improved by increasing the number of gauge configurations $N$, 
but this becomes exponentially expensive at large separations.
Maintaining a constant relative error requires $N \propto \mathrm{exp}\left(2E~|x_0-y_0|\right)$.

It is therefore essential to both develop algorithms that mitigate the exponential signal-to-noise degradation and design operators with improved overlap to the physical states.
For the former, multilevel algorithms offer a powerful way to reduce the exponential signal-to-noise degradation by exploiting the locality of the action and of the observables. In particular, Refs.~\cite{Barca:2024fpc, Barca:2023arw} show that, for pure-gauge observables, two-level sampling achieves a variance scaling as $1/N_1^2$ (with $N_1$ the number of second-level submeasurements), up to boundary effects that depend exponentially on the distance of the operators from the frozen regions.
 
As for the latter, quark-field smearing techniques such as \emph{distillation} \cite{Peardon:2009gh} can be combined with link smearing methods \cite{APE:1987ehd} to reduce excited-state contamination.
A variational basis can then be constructed to solve a Generalised Eigenvalue Problem (GEVP)
and use the resulting eigenvectors to project the operators toward the desired states
\cite{Blossier:2009kd, Barca:2022uhi, Barca:2024hrl}.

In this work we employ standard distillation, both to enhance quark-field smearing and 
to efficiently handle the increasing number of Wick contractions relevant for 
future multi-particle analyses.
Within the distillation framework \cite{Peardon:2009gh},
the disconnected contractions in Eq.~\eqref{corrql} can be expressed as
\begin{align}
\label{corrqldist}
&\langle \mathrm{O}_\Gamma(x_0) \bar{\mathrm{O}}_{\Gamma'}(y_0)\rangle_{F, \rm disc} 
=
\mathrm{Tr} \left[\tau(x_0, x_0) \Gamma \right] ~\mathrm{Tr}\left[\tau(y_0,y_0) \Gamma'\right]\,,
\end{align}
where $\tau$ are the perambulators, whose expressions are
\begin{align}
\label{peramb}
\tau_{nm}(z_0,z_0)_{\alpha \beta}&=v^{\alpha}_n(z_0)^\dagger D^{-1} ~ v^{\beta}_m(z_0)\,,
\end{align}
with $z_0=x_0,y_0$. Here $v_n$ denotes the $n$th eigenvector of the three-dimensional Laplacian operator
\begin{equation}
\nabla^2(t)_{\vec{x}, \vec{y}} = 
-6 \delta_{\vec{x}, \vec{y}} 
~
+ \sum_{k=1}^{3} 
\Bigl[
U_k(\vec{x}, t) \delta_{\vec{x} + \hat{k}, \vec{y}}
+
U^\dagger_k(\vec{x}-\hat{k}, t) \delta_{\vec{x} - \hat{k}, \vec{y}}
\Bigr]\,,
\end{equation}
constructed from the gauge fields $U$.
Each eigenvector is used as a spatial source for all Dirac components, so that the same set of spatial eigenvectors is employed for the inversions in every spin channel.
This can be viewed as an embedding of the eigenvectors to the Dirac subspace.
In Eq.~\eqref{corrqldist}, the sum over the eigenvector indices $n,m = 1,\ldots,N_v$ is implied.
By truncating the sum to a small $N_v$, the quark fields are effectively smeared, producing extended interpolating operators built from the low modes of the Laplacian operator.

\section{Two-level sampling for fermionic observables in quenched QCD}
\label{section3}

\subsection{Factorisation of the quark propagator}
In the quenched approximation, the two-point functions in Eq.~\eqref{corr} depend on the pure-gauge action $S_g[U]$ and on the Wick contractions
$\langle \mathrm{O}(t_1)\,\bar{\mathrm{O}}(t_0)\rangle_F$.
The action $S_g[U]$ is local in the gauge fields, being constructed from Wilson plaquettes.

For fermionic observables, the Wick contractions are traces of products of quark propagators (see, e.g., Eq.~\eqref{corrql}), and each propagator depends on the gauge fields over the full space–time.
A factorisation of the propagator, and hence of the fermionic correlation function, into domain-local pieces, first developed in Ref.~\cite{Ce:2016idq}, is therefore required for multilevel sampling.

\begin{figure}[t!]
\includegraphics[width=0.9\textwidth]{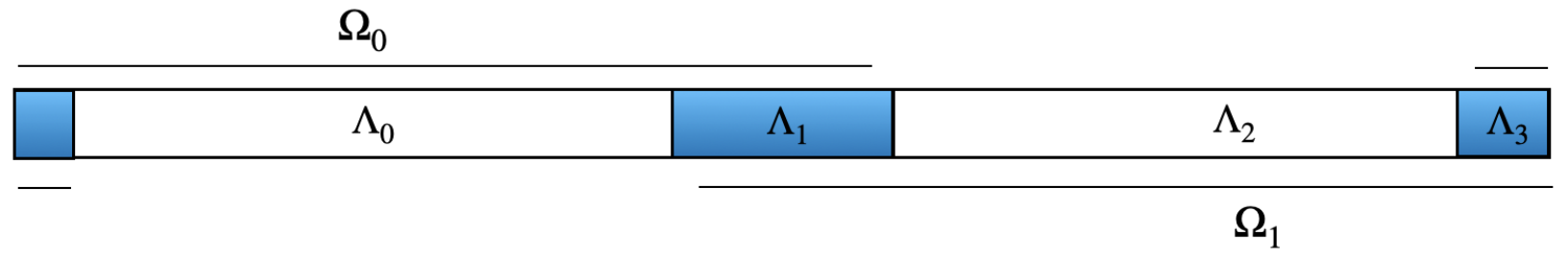}
\caption{Domain decomposition of the temporal lattice extent used in this analysis. 
The temporal extent is factorised in four regions:  $\Lambda=\Lambda_0 \oplus \Lambda_1 \oplus \Lambda_2 \oplus \Lambda_3$, 
and we consider the two overlapping domains $\Omega_0= \Lambda_3 \oplus \Lambda_0 \oplus \Lambda_1$, 
and $\Omega_1= \Lambda_1 \oplus \Lambda_2 \oplus \Lambda_3$.}
\label{Figure:dd}
\end{figure}

We decompose the temporal extent of the lattice $\Lambda$ into four regions,
$\Lambda = \Lambda_0 \oplus \Lambda_1 \oplus \Lambda_2 \oplus \Lambda_3$ (Fig.~\ref{Figure:dd}),
and introduce the overlapping domains
$\Omega_0 = \Lambda_3 \oplus \Lambda_0 \oplus \Lambda_1$ and
$\Omega_1 = \Lambda_1 \oplus \Lambda_2 \oplus \Lambda_3$.

With this domain decomposition, the factorised expression of a quark propagator travelling from $x$ to $x$ with $x_0\in\Lambda_2$ reads~\cite{Giusti:2017ksp, Ce:2016idq, Ce:2016qto}
\begin{equation}
\label{fact_qp_x_0}
D^{-1}(x,x) \;=\;
D^{-1}_{\Omega_1}(x,x) \;+\;
D^{-1}_{\Omega_1}(x,\sbullet)\,D_{\Lambda_{b0}}\,D^{-1}_{\Omega_0}\,D_{\Lambda_{b2}}\,(1-w)^{-1}\,D^{-1}_{\Omega_1}(\sbullet, x)\,,
\end{equation}
where $\sbullet$ represents all intermediate points, while $D_{\Lambda_{b0}}$ and $D_{\Lambda_{b2}}$ denote the hopping terms from all frozen regions to the active regions $\Lambda_0$ and $\Lambda_2$, and
\begin{equation}
w = PD^{-1}_{\Omega_1}\,D_{\Lambda_{b0}}\,D^{-1}_{\Omega_0}\,D_{\Lambda_{b2}}\,
\end{equation}
with $P$ being the projector to the surface of the boundaries.
Employing the Neumann series,
\begin{equation}
\frac{1}{1-w} \;=\; \sum_{n=0}^{\infty} w^n\,,
\end{equation}
which converges exponentially with $n$, yields
\begin{equation}
\label{fact_qp_x}
D^{-1}(x,x) \;=\;
D^{-1}_{\Omega_1}(x,x) \;+\;
D^{-1}_{\Omega_1}(x,\sbullet)\,D_{\Lambda_{b0}}\,D^{-1}_{\Omega_0}\,D_{\Lambda_{b2}}
\sum_{n=0}^{\infty}\!\left(PD^{-1}_{\Omega_1}\,D_{\Lambda_{b0}}\,D^{-1}_{\Omega_0}\,D_{\Lambda_{b2}}\right)^{\!n}
D^{-1}_{\Omega_1}(\sbullet, x)\,.
\end{equation}
A similar expression follows for the propagator from $y$ to $y$ with $y_0\in\Lambda_0$.
The first term in Eq.~\eqref{fact_qp_x} describes propagation confined to $\Omega_1$; the remaining terms account for propagation into $\Omega_0$ and back.
These contributions are exponentially suppressed at large $n$.
Heuristically~\cite{Parisi:1983ae, Luscher:2010ae}, on a given gauge field
\begin{equation}
\label{quark_prop_exp}
\bigl\|D^{-1}(y,x)\bigr\| \;\sim\; \exp\!\left(-\tfrac{1}{2}m_\pi\,|y-x|\right)\,,
\end{equation}
with the norm $\|A\|=\sqrt{\mathrm{Tr}\{A\,A^\dagger\}}$, so contributions mediated by distant regions decay with the distance.
Accordingly, the norm of the $n$-th term in the sum of Eq.~\eqref{fact_qp_x} is suppressed as $\exp\left(-m_\pi\,\Delta\,n\right)$, where $\Delta$ is the thickness of the overlaps $\Lambda_1,\Lambda_3$.

Using Eq.~\eqref{fact_qp_x}, a two-level integration can be applied to the two-point function in Eq.~\eqref{corrql}.
Retaining only the leading (fully local) term in Eq.~\eqref{fact_qp_x}, 
the two-level estimator for this first approximation to the isosinglet single-meson two-point functions with $\Gamma=\Gamma'$ reads
\begin{equation}\label{corr2lvl}
C^{\mathrm{2lvl}}_{\Gamma,(0)}(x_0-y_0)
=
\frac{1}{\mathcal{Z}_B}\!\int [dU_B]\, e^{-S_g[U_B]}\;
\bigl[\mathcal{O}^{\Omega_1}_\Gamma(x_0)\bigr]\;\bigl[\bar{\mathcal{O}}^{\Omega_0}_{\Gamma}(y_0)\bigr]\,,
\end{equation}
where the subscript $(0)$ denotes the first term in the factorisation, and the local (second-level) averages of the quark loops are
\begin{equation}\label{ql_omega1}
\bigl[\mathcal{O}^{\Omega_1}_{\Gamma}(x_0)\bigr]
=
\frac{1}{\mathcal{Z}_g[U_1]}\!\int [dU_1]\,\sum_{\vec{x}}\,
\mathrm{Tr}\!\left[D^{-1}_{\Omega_1}(x,x)\,\Gamma\right]\,,
\end{equation}
with an analogous expression for $\bigl[\bar{\mathcal{O}}^{\Omega_0}_{\Gamma}(y_0)\bigr]$.
If both loops reside in the same domain $\Omega_r$, a two-level separation is not possible and the standard (global) average must be used:
\begin{equation}\label{corr1lvl}
C^{\mathrm{1lvl}}_{\Gamma,(0)}(x_0-y_0)
=
\frac{1}{\mathcal{Z}_g}\!\int [dU]\, e^{-S_g[U]}\;
\mathcal{O}^{\Omega_r}_\Gamma(x_0)\,\bar{\mathcal{O}}^{\Omega_r}_{\Gamma}(y_0)\,.
\end{equation}
This one-level estimator is also used for global corrections (see below).

The statistical behaviour differs between the two schemes when observables are estimated with Markov-chain Monte Carlo (e.g., HMC).
While the variance of the standard estimator scales as $1/N_1$, the leading contribution to the two-level variance scales as $1/N_1^2$ (up to exponential boundary terms), mirroring the pure-gauge case~\cite{Barca:2024fpc}.

A key difference relative to pure-gauge observables is that $D^{-1}_{\Omega_1}(x,x)$ is only an approximation to the full propagator $D^{-1}(x,x)$.
To correct for this, we also compute the standard two-point function with the solver on the full lattice,
\begin{equation}\label{corr1lvl_full}
C^{\mathrm{1lvl}}_{\Gamma}(x_0-y_0)
=
\frac{1}{\mathcal{Z}_g}\!\int [dU]\, e^{-S_g[U]}\;
\mathcal{O}_\Gamma(x_0)\,\bar{\mathcal{O}}_\Gamma(y_0)\,,
\end{equation}
and define the leading correction
\begin{equation}
\label{correction}
\delta^{(0)}_\Gamma(x_0,y_0)
=
C^{\mathrm{1lvl}}_\Gamma(x_0,y_0) - C^{\mathrm{1lvl}}_{\Gamma,(0)}(x_0,y_0)\,.
\end{equation}
The improved two-level estimator is then
\begin{equation}
\label{c2pt_2lvl_corrected}
C^{\mathrm{2lvl}}_\Gamma(x_0,y_0)
=
C^{\mathrm{2lvl}}_{\Gamma,(0)}(x_0,y_0) + \delta^{(0)}_\Gamma(x_0,y_0)\,.
\end{equation}
Beyond the leading contribution, we also compute the next term in the Neumann series in Eq.~\eqref{fact_qp_x}, which provides a second-order approximation to the propagator. This first correction accounts for propagation from one active region into the other and back, and can itself be treated with a two-level integration (see Sec.~\ref{sec:sec3b}). Including this contribution improves the accuracy of the approximation near the frozen boundaries while preserving the variance reduction of the two-level scheme. Higher-order terms can likewise be evaluated and would further extend the two-level error reduction in regimes where $\delta^{(n)}_\Gamma(x_0, y_0)$, corresponding to the globally added one-level correction to the $n$-th approximation of the quark propagator, begins to dominate.

Alternatively, a multiboson approximation~\cite{Luscher:1993xx} could be used to treat all residual corrections; we find the direct correction via the full solver already efficient and do not pursue the multiboson approach here.
\subsection{Analysis of disconnected two-points functions with one- and two-level estimators}
\label{sec:sec3b}
The procedure to generate the gauge configurations follows the pure-gauge setup of Ref.~\cite{Barca:2024fpc}.
First, $N_0$ gauge configurations are generated by updating the gauge fields over the full space–time; these are the “boundary” fields $U_B$ that appear in Eq.~\eqref{corr2lvl}.
Second, for each stored $U_B$ we generate a sequence of $N_1$ gauge fields updated only in the active temporal regions $\Lambda_0$ and $\Lambda_2$, while $\Lambda_1$ and $\Lambda_3$ are frozen.
In total, $N_0\times N_1$ configurations are produced.

On these configurations we compute $N_v$ Laplacian eigenvectors and estimate traces of quark loops using the full and the domain-restricted Wilson–Dirac propagators:
\begin{align}
\label{1pt_std}
\mathcal{O}_{\Gamma, ij}(x_0) &= \sum_{n=1}^{N_v}\mathrm{Tr}\left[v_n^\dagger(x_0) D^{-1}_{ij} v_n(x_0) \Gamma \right]\,,
\\
\label{1pt_approx_omega1}
\mathcal{O}^{\Omega_r}_{\Gamma, ij}(x_0) &= \sum_{n=1}^{N_v}\mathrm{Tr}\left[v_n^\dagger(x_0) D^{-1}_{\Omega_r, ij} v_n(x_0) \Gamma \right]
\,,
\end{align}
with $D^{-1}_{\Omega_r} = D^{-1}_{\Omega_0}$ for $x_0\in \Omega_0$ and $D^{-1}_{\Omega_r} = D^{-1}_{\Omega_1}$ for $x_0\in \Omega_1$. 
Here $i=1,\ldots,N_0$ and $j=1,\ldots,N_1$ label the gauge fields $U_{ij}$.
The disconnected two-point function with the full propagator (standard sampling) is
\begin{equation}
\label{c2pt_std}
C^{\textnormal{1lvl}}_{\Gamma}(x_0, y_0)
=
\frac{1}{N_0 N_1}\sum_{ij} \mathcal{O}_{\Gamma, ij}(x_0) \bar{\mathcal{O}}_{\Gamma, ij}(y_0)\,.
\end{equation}
Using the approximated propagators instead, the one- and two-level estimators corresponding to Eqs.~\eqref{corr2lvl} and \eqref{corr1lvl} read
\begin{align}
\label{c2pt_1lvl_approx}
C^{\textnormal{1lvl}}_{\Gamma, (0)}(x_0, y_0)
&=
\frac{1}{N_0 N_1}\sum_{ij} \mathcal{O}^{\Omega_1}_{\Gamma, ij}(x_0) \bar{\mathcal{O}}^{\Omega_0}_{\Gamma, ij}(y_0)\,,
\\
\label{c2pt_2lvl_approx}
{C}^{\textnormal{2lvl}}_{\Gamma, (0)}(x_0, y_0)
&=
\frac{1}{N_0}\sum_{i}
\left[\mathcal{O}^{\Omega_1}_{\Gamma, i}(x_0)\right]
\left[\bar{\mathcal{O}}^{\Omega_0}_{\Gamma, i}(y_0)\right]\,.
\end{align}
Here the square brackets denote second-level averages in the active regions (cf. Eqs.~\eqref{corr2lvl}–\eqref{ql_omega1}), 
i.e. for the quark loops in $\Omega_1$,
\begin{equation}
\label{first_approx_loops}
\left[\mathcal{O}^{\Omega_1}_{\Gamma, i}(x_0)\right]
=
\frac{1}{N_1} \sum_{j=1}^{N_1} \mathcal{O}^{\Omega_1}_{\Gamma, ij}(x_0)\,.
\end{equation}
The only difference between Eqs.~\eqref{c2pt_std} and \eqref{c2pt_1lvl_approx} is the propagator approximation; 
the corresponding correction is
\begin{equation}
\label{correction}
\delta^{(0)}_\Gamma(x_0, y_0) = C^{\textnormal{1lvl}}_{\Gamma}(x_0, y_0) - {C}_{\Gamma, (0)}^{\textnormal{1lvl}}(x_0, y_0)\,.
\end{equation}
Adding this to the two-level estimator of the first approximation gives the corrected result
\begin{equation}
\label{c2pt_2lvl_corrected}
C^{\textnormal{2lvl}}_{\Gamma}(x_0, y_0) = C_{\Gamma, (0)}^{\textnormal{2lvl}}(x_0, y_0) + \delta^{(0)}_\Gamma(x_0, y_0)\,.
\end{equation}
For the two-level method to be efficient, the fluctuations of $\delta^{(0)}_\Gamma$ (computed using one-level sampling) must remain much smaller than those of $C^{\textnormal{2lvl}}_{\Gamma,(0)}$; otherwise the correction hinders the two-level gain.
Since the norm of the $n$th term in Eq.~\eqref{fact_qp_x} is expected to be suppressed as $\exp(-m_\pi n\Delta)$, the onset of saturation can be anticipated from the thickness $\Delta$ of the frozen regions and $N_1$.
In order to improve the approximation and the statistical precision, additional terms of the Neumann series in Eq.~\eqref{fact_qp_x} can be included in the two-level treatment.

Separating the $n=0$ term, the first correction to the fully factorised loop is
\begin{align}
\label{firstcorr_ql_omega1}
\mathcal{O}^{(1)}_{\Gamma}(x_0) &= 
\sum_{n=1}^{N_v} \mathrm{Tr}\left[v_n^\dagger D^{-1}_{\Omega_1} D_{\Lambda_{b0}}D^{-1}_{\Omega_0}D_{\Lambda_{b2}}D^{-1}_{\Omega_1}v_n\Gamma \right]\,,
\\
\label{firstcorr_ql_omega0}
\mathcal{O}^{(1)}_{\Gamma}(y_0) &= 
\sum_{n=1}^{N_v} \mathrm{Tr}\left[v_n^\dagger D^{-1}_{\Omega_0} D_{\Lambda_{b2}}D^{-1}_{\Omega_1}D_{\Lambda_{b0}}D^{-1}_{\Omega_0}v_n\Gamma \right]\,,
\end{align}
for $x_0\in\Omega_1$, $y_0\in\Omega_0$ and for $b$ we implicitly include contributions from the frozen regions.
The corresponding first-order correction to the two-point function is estimated as
\begin{align}
\label{firstcorr_2pt}
C^{\rm 1lvl}_{\Gamma, (1)}(x_0, y_0) = 
\frac{1}{N_0 N_1}\sum_{ij} 
\left[
\mathcal{O}^{(1)}_{\Gamma, ij}(x_0) \bar{\mathcal{O}}^{\Omega_0}_{\Gamma, ij}(y_0)
~+
\mathcal{O}^{\Omega_1}_{\Gamma, ij}(x_0) \bar{\mathcal{O}}^{(1)}_{\Gamma, ij}(y_0)
~+
\mathcal{O}^{(1)}_{\Gamma, ij}(x_0) \bar{\mathcal{O}}^{(1)}_{\Gamma, ij}(y_0)
\right]\,.
\end{align}
Physically, this correction accounts for contributions where one quark loop stays in one region
and the other quark loop propagates to the other region and comes back, plus the contribution where both quark loops propagate to the other region and come back to the same region.

To perform a two-level integration, contributions supported in $\Omega_0$ and $\Omega_1$ must be separated and averaged independently.
Numerically, we first average terms in $\Omega_0$ and $\Omega_1$ over gauge updates restricted to those domains, then average over boundary fields.
To this end, we insert Laplacian eigenvectors and/or stochastic vectors at the boundary of the frozen regions to cut the fermionic lines in $\Omega_0$ and $\Omega_1$. This allows the second-level and independent sampling of the propagators in these regions. 
In particular, we approximate the identity via
\begin{equation}
\label{approx_1}
1\approx P_v + P_{\perp} \equiv\sum_{k=1}^{N_{\tilde{\eta}}} \tilde{\eta}_k (\tilde{\eta}_k)^\dagger\,,
\end{equation}
where
\begin{equation}
P_v = \sum_{\beta=0}^{3}\sum_{v=1}^{N_v^b} v_i^\beta (v_i^\beta)^\dagger\,,
\qquad
P_\perp = \frac{1}{N_{\eta^\perp}}\sum_{k=1} ^{N_{\eta^\perp}} \eta^\perp_k (\eta_k^\perp)^\dagger\,.
\end{equation}
In these expressions, $P_v$ is the low-mode Laplacian projector, already used within distillation to compute smeared interpolating operators at source and sink, 
and $P_\perp$ is used to estimate stochastically the remaining part of the identity matrix.
The latter is done by introducing stochastic vectors $\eta_k$, in this work $Z_2$ noise vectors, and computing
\begin{equation}
\label{stoch_vec}
\eta^{\perp}_k = \left(1 -\mathbb{P}_v \right)\,\eta_k \,.
\end{equation}
Hence, the total number of \textit{deflated} vectors in Eq.~\eqref{approx_1} are $N_{\tilde{\eta}}=4N_{v}^b + N_{\eta^\perp}$.
Notice that the number of eigenvectors used at the boundary of the frozen regions ($N^b_v$) can differ from the number of eigenvectors used to smear the interpolating operator ($N_v$). Choosing $N_v^b\le N_v$ allows reuse of solution vectors $D^{-1}_{\Omega_r}v_n$ in both the leading term and the first correction.

For each \textit{deflated} stochastic vector $\tilde{\eta}_k$, located at the time slice of the boundary, we compute
\begin{equation}
\label{firstcorr_ql_omega1_eta}
\mathcal{O}^{(1)}_{\Gamma, k}(x_0) = \sum_{n=1}^{N_v}\mathrm{Tr}\left\{v_n^\dagger D^{-1}_{\Omega_1} D_{\Lambda_{10}} D^{-1}_{\Omega_0}D_{\Lambda_{12}} \tilde{\eta}_k \tilde{\eta}_k^\dagger D^{-1}_{\Omega_1}v_n\Gamma \right\}\,,
\end{equation}
and the analogous expression for $y_0 \in \Lambda_2$.
As $N_{\tilde{\eta}}\!\to\!\infty$ this reproduces Eqs.~\eqref{firstcorr_ql_omega1}–\eqref{firstcorr_ql_omega0}.
This stochastic representation enables a two-level average of Eq.~\eqref{firstcorr_2pt}.
Using cyclicity of the trace, the two-level estimator of the first term becomes
\begin{equation}
\label{firstcorr1_2pt}
\langle \mathcal{O}^{(1)}_{\Gamma}(x_0) ~\bar{\mathcal{O}}^{\Omega_0}_{\Gamma}(y_0) \rangle^{\rm 2lvl}
=
\frac{1}{N_0}
\sum_{i=1}^{N_0}
\sum_{k=1}^{N_{\tilde{\eta}}} \left[\psi^{\Omega_1}_k(x_0)^\dagger\right]_i ~ \left[\psi_k^{\Omega_0}(y_0)\right]_i\,,
\end{equation}
where
\begin{align}
&\left[\psi^{\Omega_1}_k(x_0)^\dagger\right]_i
=
\frac{1}{N_1}
\sum_{j=1}^{N_1}
\left(\tilde{\eta}_k^\dagger D^{-1}_{\Omega_1}v_n(x_0)\Gamma v_n^\dagger(x_0) D^{-1}_{\Omega_1} P_{\Lambda_{10}}\right)_{|~U_{ij}}\,,
\\
\label{psi_k_omega1}
&\left[\psi_k^{\Omega_0}(y_0)\right]_i
=
\frac{1}{N_1}
\sum_{j=1}^{N_1}
\left(D_{\Lambda_{10}}  D^{-1}_{\Omega_0}D_{\Lambda_{12}} \tilde{\eta}_k ~\bar{\mathcal{O}}^{\Omega_0}_{\Gamma}(y_0) \right)_{|~U_{ij}}\,.
\end{align}
Here $U_{ij}$ denotes the gauge field used in the evaluation, and $\bar{\mathcal{O}}^{\Omega_0}_{\Gamma}(y_0)$ is a complex number for each $U_{ij}$. Furthermore, $P_{\Lambda_{10}}$ is a projector to the boundary $\Lambda_{10}$, which is used to reduce the dimensionality of the solution vector for disk storage.

The solution vectors $\chi_n = D^{-1}_{\Omega_1} v_n(x_0)$ are reused both here and in Eq.~\eqref{1pt_approx_omega1}; the only additional inversions are those involving boundary sources, 
$D_{\Lambda_{10}}\, D^{-1}_{\Omega_0}\, D_{\Lambda_{12}}\, \tilde{\eta}_k$.  
These boundary inversions are independent of the specific source and sink positions $x_0$ and $y_0$, so they can be computed once and reused throughout.  
Placing the deflated vectors next to $D^{-1}_{\Omega_0}$ in Eq.~\eqref{firstcorr_ql_omega1_eta} permits recycling of the low-mode solutions $D^{-1}_{\Omega_r} v_n$.

Similar expressions for the two-level estimators of the second and third terms in Eq.~\eqref{firstcorr_2pt}.
The third term in Eq.~\eqref{firstcorr_2pt} is analogous and is expected to be exponentially suppressed relative to the first two contributions, since it contains first-order corrections at both source and sink. We confirm this numerically and discuss it in the next section.

In particular, in the next section we present numerical results for the scalar ($\Gamma=I$) one-point functions from Eq.~\eqref{1pt_approx_omega1} and compare them to the unbiased one-level (full-solver) results in Eq.~\eqref{1pt_std}, thereby benchmarking the propagator factorisation.
We then compare standard and two-level estimates of the two-point functions for several choices of $\Gamma$.

Two important differences with respect to the pure-gauge study are:
\begin{itemize}
    \item The two-level algorithm is applied to the leading and/or next-to-leading terms of the propagator factorisation, with the residual global correction added from the one-level estimator;
    \item The thickness $\Delta$ of the frozen regions is larger than in the pure-gauge case.
    The increased thickness leads to a suppression of higher-order terms in the convergence series of the propagator and hence correlation functions. In addition, a sufficiently thick frozen region is required in the multiboson factorisation of the fermion determinant in dynamical QCD studies to ensure rapid convergence of the associated Neumann series~\cite{Ce:2016ajy, Giusti:2017ksp, Ce:2017ndt}.
\end{itemize}
As a consequence of the increased thickness, however, the two-level error reduction at fixed separation $|x_0-y_0|$ is degraded:
the two-level variance saturates at the level of fluctuations propagating from both the boundaries, which do not scale with $1/N_1$~\cite{Barca:2024fpc}; when the frozen regions are thicker, the effective distance to those boundaries is smaller (at fixed $|x_0-y_0|$), and saturation sets in earlier.
A central goal of this study is to determine whether, despite these effects, two-level sampling remains more efficient than standard sampling for the fermionic observables considered here.

\section{Numerical results}\label{numericalres}

In this section we present our numerical results and address several key questions regarding the performance of the factorised two-point function and its corrections:
\begin{itemize}
    \item Does the statistical error of the approximated two-point function in Eq.~\eqref{c2pt_2lvl_approx} decrease as $1/N_1$ with the number of level-1 updates?
    \item Once the correction is included using standard sampling, does the error of the full two-point function in Eq.~\eqref{c2pt_2lvl_corrected} still follow a $1/N_1$ scaling?
    \item For which values of $N_1$ does the correction term obtained from standard sampling become significant and start to dominate the total error?
    \item Is it sufficient to evaluate only the leading correction in Eq.~\eqref{firstcorr_2pt} with a two-level sampling (e.g.~ via Eq.~\eqref{firstcorr1_2pt} for the first term) to further reduce the statistical uncertainty?
    \item Given that larger frozen regions are expected to impact two-level efficiency, do we obtain reliable signals for the effective masses?
\end{itemize}
In the following subsections we describe the simulation setup and present results addressing these raised questions.

\subsection{Details of the simulations}
We simulate the four-dimensional $\mathrm{SU}(3)$ Yang–Mills theory using the Wilson gauge action with periodic boundary conditions.
Gauge configurations are generated at $\beta=6.0$, corresponding to $a=0.0876~\mathrm{fm}$~\cite{Necco:2001xg}, on a volume $V/a^4=64\times 16^3$.
To perform the two-level integration, we use molecular dynamics updates as in Ref.~\cite{Barca:2024fpc}.
In the simulation, the frozen regions are $t/a\in\{0,1,61,62,63\}$ for $\Lambda_3$ and $t/a\in\{29,30,31,32,33\}$ for $\Lambda_1$, while the remaining time slices constitute $\Lambda_0$ and $\Lambda_2$.

We generate $N_0=101$ level-0 (boundary) configurations and for each of them a sequence of $N_1=1000$ level-1 (local) updates; the resulting fields are denoted $U_{ij}$ with $i=1,\ldots,N_0$ and $j=1,\ldots,N_1$.
On each $U_{ij}$ we compute the full and the domain-restricted Wilson–Dirac propagators using a clover-improved Wilson action with $\kappa=0.13393$ and $c_{\rm SW}=1.769231$, for which the lightest pseudoscalar mass is found at $m_\pi\approx760~\mathrm{MeV}$.
The mass is tuned so that the lowest non-interacting $\pi\pi$ energy lies close to the pure-gauge scalar glueball mass (at a similar coupling, $\beta=5.99$, one finds $m_G^{0^{++}}\approx1560~\mathrm{MeV}$~\cite{Athenodorou:2020ani}).
We use $N_v=10$ Laplacian eigenvectors to smear the quark fields in the meson interpolating operators and invert the Dirac operator on these eigenvector sources.

To compute the approximated propagators we adopt a modified version of \texttt{openQCD}~\cite{openQCD, Luscher:2012av} imposing Dirichlet boundary conditions at the boundaries of the frozen regions.
For instance, $D^{-1}_{\Omega_0,ij}(y_0,y_0)$ is computed on $U_{ij}$ with inversions restricted to $\Omega_0$, thereby neglecting contributions from $\Lambda_2$ (cf. Fig.~\ref{Figure:dd}).

\subsection{One-point functions}
For a first illustration of the convergence of the Neumann series adopted for the quark propagator, we compute scalar quark loops on the full set of $N_0\times N_1$ configurations using the full propagator and the domain-restricted propagators (Eqs.~\eqref{1pt_std}–\eqref{1pt_approx_omega1}).

\begin{figure}[t!]
\centering
\includegraphics[width=0.92\textwidth]{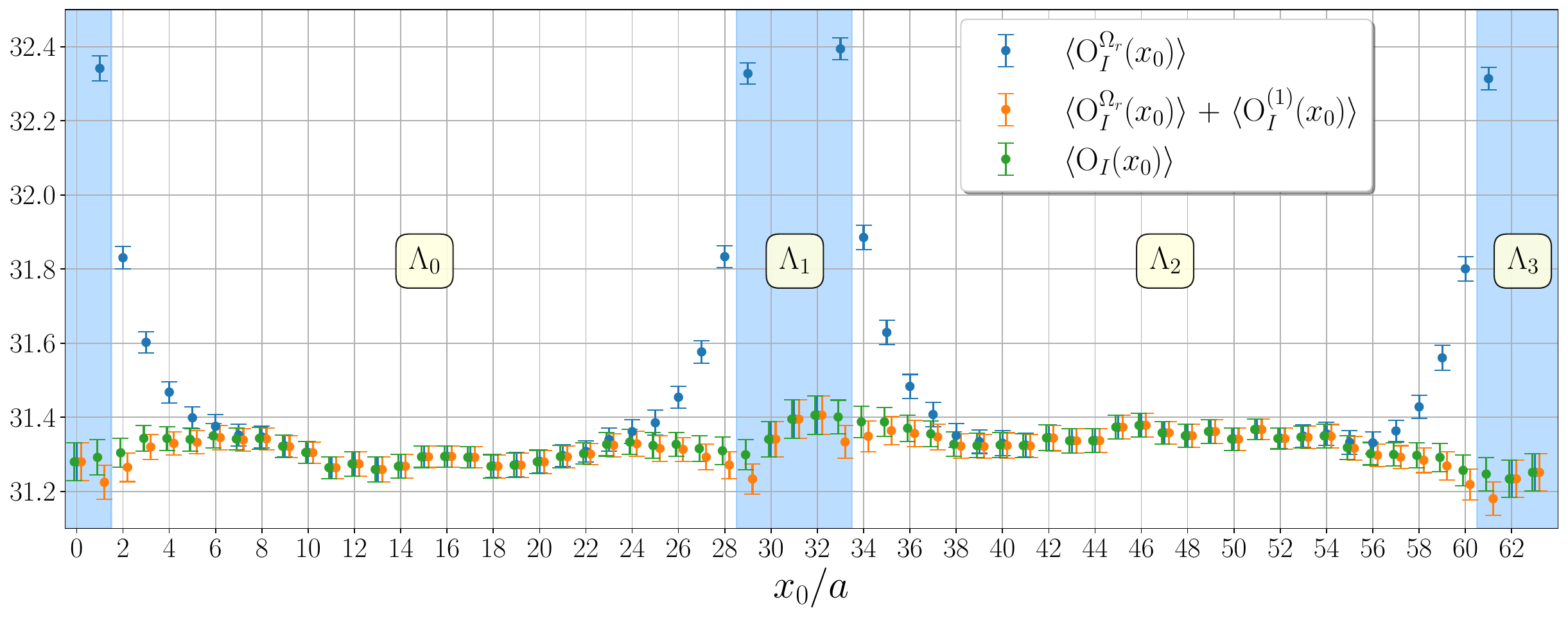} \\[1ex]
\caption{Scalar one-point functions obtained with the full propagator (green), the leading approximation (blue), 
and the leading approximation plus first correction (orange).
Blue vertical bands mark the frozen regions $\Lambda_1$ and $\Lambda_3$.
Estimates follow Eqs.~\eqref{1pt_std}–\eqref{1pt_approx_omega1}.}
\label{Figure:combined_scalar_ql}
\end{figure}

We also evaluate the first correction with standard sampling (Eqs.~\eqref{firstcorr_ql_omega1}–\eqref{firstcorr_ql_omega0}), which is then used to improve the approximation to the second order.
Fig.~\ref{Figure:combined_scalar_ql} compares the scalar one-point functions (which have non-vanishing VEVs) obtained with the full propagator (green) and with the first (blue) and second (orange) approximations; orange points are x-shifted for visibility.
The full-propagator loops fluctuate around a constant, as expected.
The first approximation to the scalar quark loop (cf. Eq.~\eqref{first_approx_loops} with $\Gamma=I$) agrees with the full result when the loop is sufficiently far from the other active region, but deviations appear as the loop approaches the boundaries.
Data at $x_0/a\in\{0,30,31,32,62,63\}$ are always computed with the full propagator (hence identical in the comparison) and are noisier because the gauge links on those time slices are updated less than in the dynamical regions; 
this difference is more visible as $N_1$ increases.
Including the first correction leads to a significant improvement of the approximation towards the exact loop near the boundaries, consistent with Eq.~\eqref{quark_prop_exp}.
In particular, within our level of accuracy, the results obtained with the second-order approximation are consistent, with the full scalar quark loops. Similar findings are obtained for the two-point functions, as discussed in the next section.
The remaining effects will be taken into account in the final estimates of the two-point functions.

\vspace{-0.7cm}

\subsection{Two-point functions}
Using Eqs.~\eqref{c2pt_std}–\eqref{c2pt_2lvl_approx} with $N_0=101$ and $N_1$ values up to $1000$ for selected time slices, 
we compare one- and two-level estimates of the disconnected two-point functions.
Our goals are: (i) verify the expected $1/N_1^2$ variance scaling of the \emph{approximated} two-level estimator (versus $1/N_1$ for one-level), and (ii) study the dependence on the distance of the operators from the frozen regions, as in the pure-gauge analysis~\cite{Barca:2024fpc}.
Because Eqs.~\eqref{c2pt_1lvl_approx}–\eqref{c2pt_2lvl_approx} are leading-order in the factorisation, we compute the global difference in Eq.~\eqref{correction} with one-level sampling and add it to the two-level result (Eq.~\eqref{c2pt_2lvl_corrected}).

\begin{figure}[t]
    \centering
    \includegraphics[width=\textwidth]{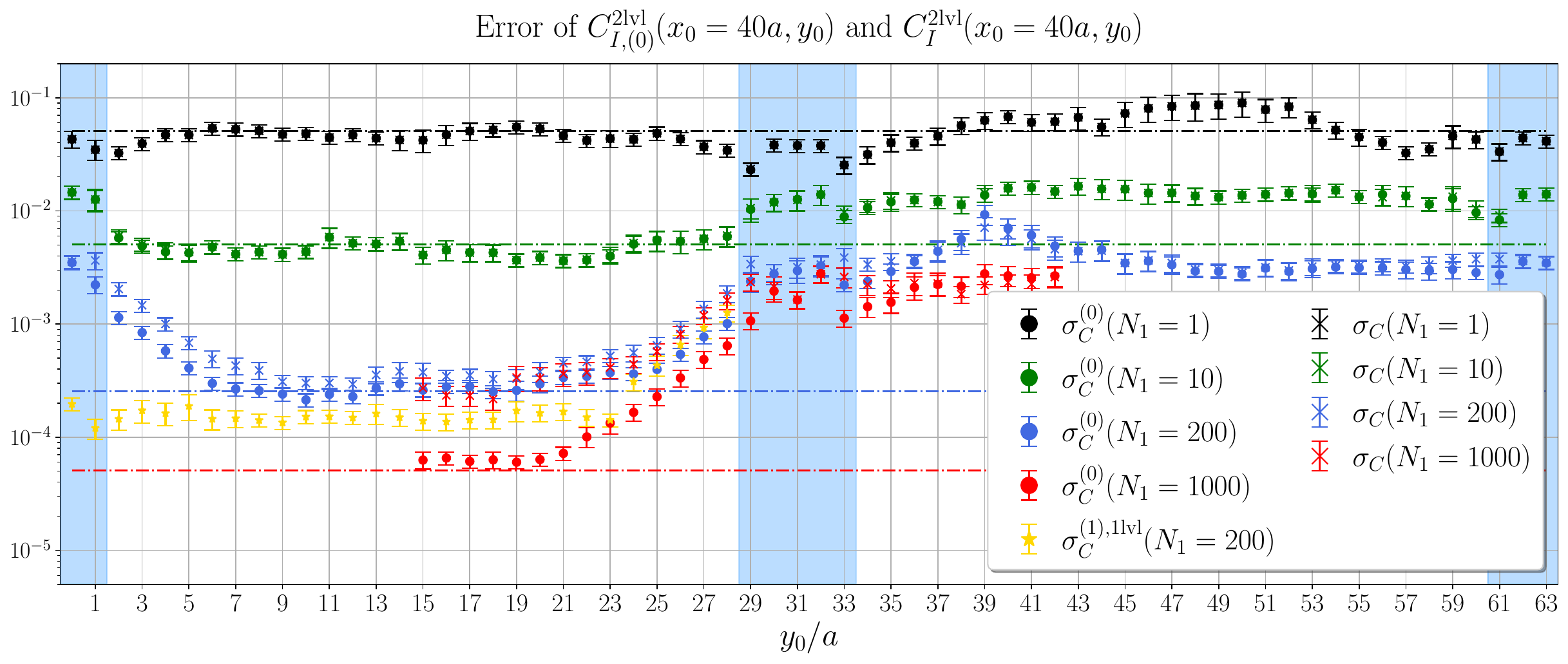}
    \caption{Statistical errors of the two-level estimators for the disconnected scalar two-point function before (open circles) and after (filled circles) adding the global correction of Equation~\eqref{correction}.
    Open (filled) circles correspond to Equations~\eqref{c2pt_2lvl_approx} (\eqref{c2pt_2lvl_corrected}).
    Blue vertical bands indicate the frozen regions $\Lambda_1$ and $\Lambda_3$.
    One loop is fixed at $t_0=40a$ while the other is varied.
    Dashed lines show ideal $1/N_1$ error scaling normalised to $\sigma_C^{(0)}(N_1\!=\!1)$ for $N_1=1,10,200,1000$.
    Yellow points: statistical error of the one-level estimate of the first correction (Eq.~\eqref{firstcorr_2pt}) with $N_0=101$, $N_1=200$.}
    \label{Figure:combined_scalar_2pt}
\end{figure}

Fig.~\ref{Figure:combined_scalar_2pt} displays the error for the scalar disconnected correlator $C_I(x_0=40a, y_0)$ from the approximated and corrected two-level estimators at different values of $N_1=1,~10,~200,~1000$.
The red data crosses show a jump at $y_0=15a$ for $N_1=1000$ because we perform $N_1=1000$ submeasurements only for $15\le y_0/a \le 42$ and $N_1=200$ for the other time slices.
When both loops lie in the same domain, the error is approximately constant in $y_0$ and decreases as $1/\sqrt{N_1}$; when they lie in different active domains, the variance follows $1/N_1^2$ up to boundary-induced corrections that decay exponentially with the distance from the frozen regions. 
We observe clear $1/N_1^2$ scaling up to $N_1\simeq200$.
For larger $N_1$, after adding the global correction computed with $N_0=101$, $N_1=200$, the total error no longer decreases: the correction’s own uncertainty dominates.
Indeed, the first correction in Eq.~\eqref{firstcorr_2pt} (implicitly added by Eq.~\eqref{correction} along with the remaining terms) carries an error of that order (yellow points).
Consequently, if only the leading approximation is treated with the two-level scheme, the corrected error cannot be reduced below the noise level of the correction unless the global difference is recomputed with higher statistics.
A more efficient strategy is therefore to evaluate the first correction itself using a two-level approach, as discussed next.
%%%

\paragraph{Analysis with first correction}
Going beyond the leading approximation and including the second order into the multilevel scheme reduces the size of the uncertainty of the corresponding global correction.
\begin{figure}[!t]
    \centering
    \includegraphics[width=1.\textwidth]{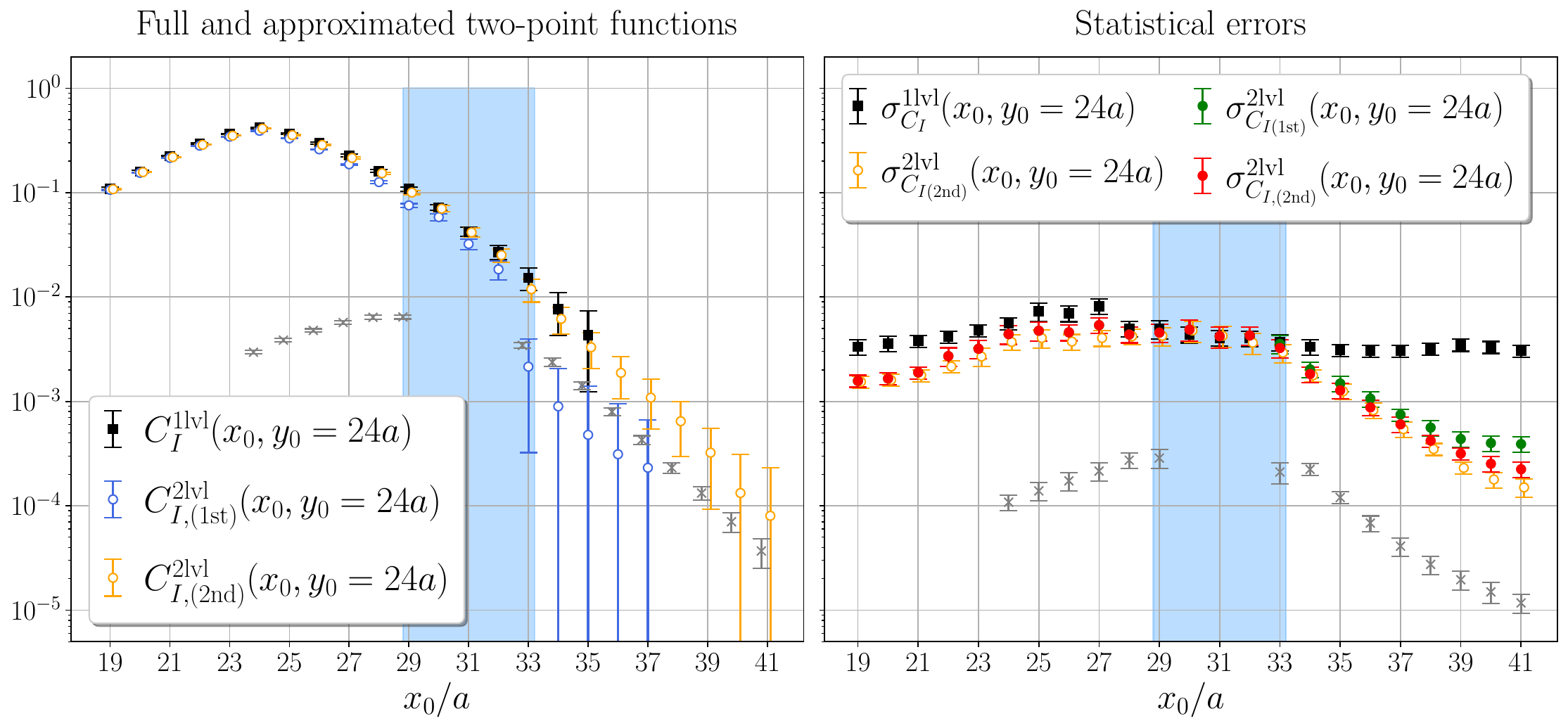}
    \caption{(left) Signals and errors of the scalar disconnected two-point functions with the full solver (black) from standard sampling, with first approximation (blue), and with second approximation (orange) of the quark propagator from two-level sampling. The source is fixed at $y_0=24a$ and the sink position $x_0$ is varied. (right) Statistical errors: empty markers are approximated results and full markers are corrected results. The grey crosses show the signal (left plot) and error (right plot) of the third term in Eq.~\eqref{firstcorr_2pt} with one-level sampling using $N_0=101$ and $N_1=200$ configurations.
    }
    \label{fig:2pts_firstcorr}    
\end{figure}
We therefore estimate the first correction to the two-point function with both one- and two-level sampling (Eqs.~\eqref{firstcorr_2pt}, \eqref{firstcorr1_2pt}) using up to $N_{\eta^\perp}=100$ stochastic vectors, deflated by $N_v^b=10$ Laplacian eigenvectors as discussed in the previous section.
Fig.~\ref{fig:2pts_firstcorr} shows the signal and errors of the two-point functions at fixed $y_0=24a$ and different $x_0$. The unbiased one-level (full-solver) signal is lost around $x_0\!\approx\!36a$ with $N_0=101$ and $N_1=200$, whereas the two-level estimators retain a clear signal at larger separations due to the exponential error reduction with boundary distance.
Adding the two-level first correction $C_{I, \rm(1)}^{\rm 2lvl}(x_0, y_0)$ to the leading approximation $C_{I, \rm(1st)}^{\rm 2lvl}(x_0, y_0) := C_{I, \rm(0)}^{\rm 2lvl}(x_0, y_0)$) yields a \emph{second} approximation $C_{I, \rm(2nd)}^{\rm 2lvl}(x_0, y_0)$) whose signal nearly coincides within the errors with the full one even near the boundaries, where higher-order terms matter most, see left plot in Fig.~\ref{fig:2pts_firstcorr}.
The residual global difference (computed with $N_1=200$ as for the first approximation) is tiny; the corrected two-level second approximation agrees within errors with the pure two-level second approximation result.

With $N_0=101$ and $N_1=200$ (one-level), the error of the third term in Eq.~\eqref{firstcorr_2pt} is about an order of magnitude smaller than the first two and its statistical uncertainty decreases exponentially with the distance from the boundaries (grey crosses in the right panel of Fig.~\ref{fig:2pts_firstcorr}); hence, we do not evaluate this term with two-level sampling.
At fixed $y_0$, we find that the variance of the second term in Eq.~\eqref{firstcorr_2pt} is approximately constant in $x_0$, while the first term decays exponentially with the distance to the boundaries.

\paragraph{Weighted average}
Since correlators at different source times $y_0$ have different variances at fixed $\Delta t=x_0-y_0$ (especially with two-level sampling), we form a weighted average~\cite{Barca:2024fpc}:
\begin{equation}
\label{weighted_average}
\bar{C}^{\rm X}(\Delta t) = \sum_{y_0} w(y_0+\Delta t, y_0) C^{\rm X}(y_0+\Delta t, y_0)\,,
\end{equation}
with $X=\textnormal{1lvl, ~2lvl}$.
The weight functions are chosen to be proportional to the inverse of the variance, 
i.e., $w(x_0, y_0) = \mathcal{N} \sigma^2(x_0, y_0)^{-1}$, 
with the normalisation $\mathcal{N}$ such that $\sum_{y_0} w(y_0+\Delta t, y_0)=1$.
\subsection{Noise-to-signal ratio}
\begin{figure}[!t]
\centering
\includegraphics[width=1.\textwidth]{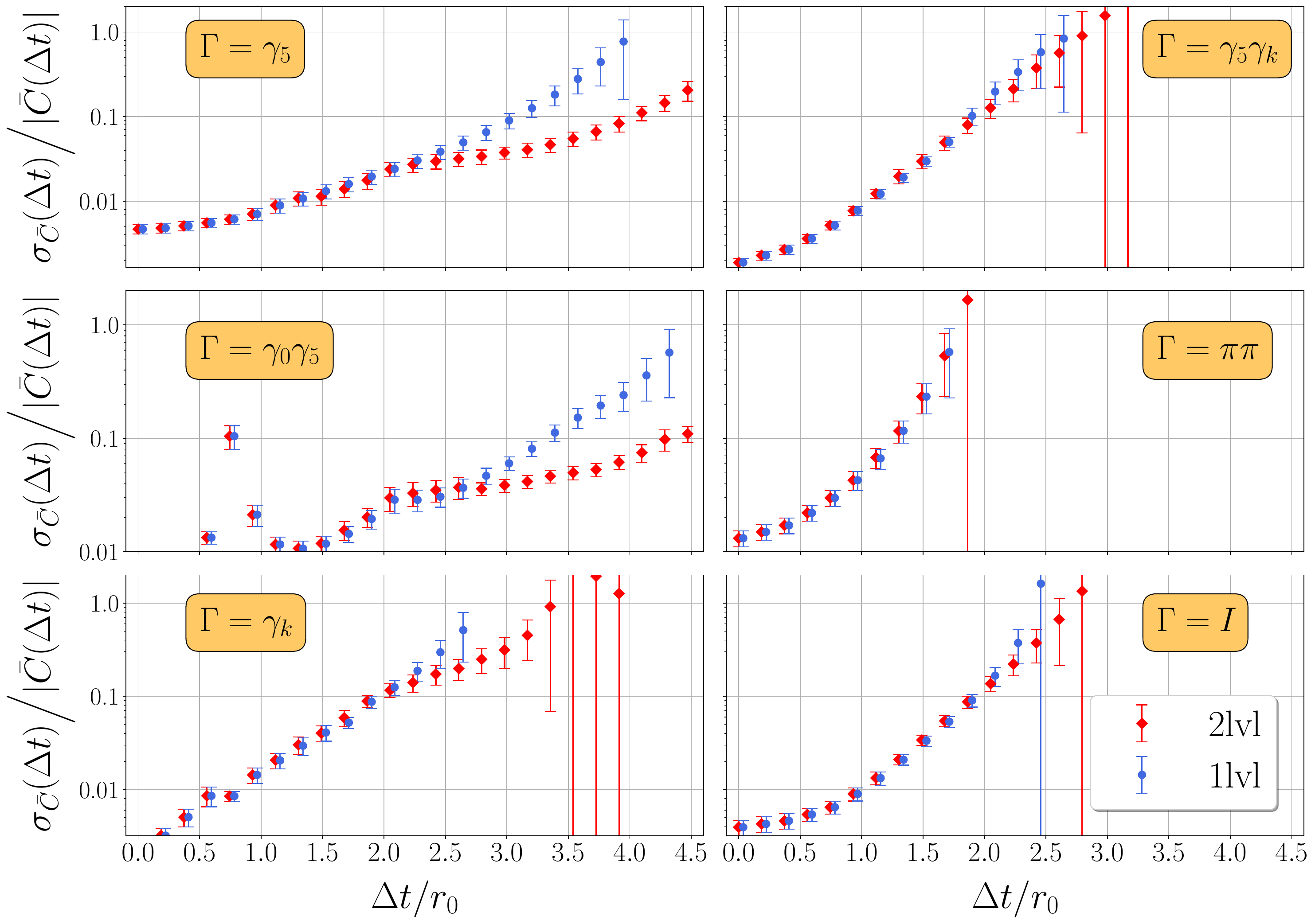}
\caption{Noise-to-signal ratios of weighted disconnected two-point functions for
$\mathrm{O}_\Gamma=\bar{\psi}\Gamma\psi$ and $\Gamma=I,\,\gamma_5,\,\gamma_\mu,\,\gamma_i\gamma_j\epsilon_{ijk},\,\gamma_5\gamma_k$.
One-level: $N_0=101$, $N_1=200$; two-level: $N_0=101$, $N_1=200$.}
\label{fig:noise-to-signal}
\end{figure}

Using the weighted correlators, we observe the following scaling for the variance:
\begin{align}
    \label{scaling1lvl}
    \frac{\sigma^2_{\bar{C}}(\Delta t)}{\bar{C}^2(\Delta t)}
    &\approx
    \frac{\tilde{a}_0^2}{N_0N_1} e^{2m^\Gamma \Delta t}
    \hspace{1.5cm} \text{for $\Delta t \lesssim \Delta \tilde{t}$~,}
    \\
    \label{scaling2lvl}
        \frac{\sigma^2_{\bar{C}}(\Delta t)}{\bar{C}^2(\Delta t)}
    &\approx
    \frac{\tilde{c}_0^2}{N_0N_1^2} e^{2m^\Gamma \Delta t}
    +
    \frac{2\tilde{c}_1^2}{N_0 N_1} e^{m^\Gamma \Delta t}
    +
    \frac{\tilde{c}_2^2}{N_0} 
    \hspace{1.5cm}
    \text{for $\Delta t \gtrsim \Delta \tilde{t}$},
\end{align}
where $m^\Gamma$ is the mass gap of the disconnected two-point functions in the channel $\Gamma$.
The noise-to-signal scaling in Eq.~\eqref{scaling1lvl} is typical with standard sampling algorithms.
Fig.~\ref{fig:noise-to-signal} shows the noise-to-signal ratios for the disconnected correlators considered here, including the $I=0$ of the $\pi\pi$ channel.

In particular, we can see the two different scalings in Eqs.~\eqref{scaling1lvl}-\eqref{scaling2lvl}: 
at short distance, the noise-to-signal ratio grows exponentially as fast as the one-level error.
After some intermediate distance $\Delta \tilde{t}$, which depends on the channel, the two-level noise-to-signal error does not grow as fast as the one-level error, thus enabling the analysis of the correlation functions for larger distances.
This scaling is similar to what observed in the pure-gauge study~\cite{Barca:2024fpc}.

However, compared to the pure-gauge study, the two-level regime sets in at larger separations (in units of $r_0$). For instance, in the channel $\Gamma=I$, the crossover distance with $N_1=1000$ and similar $N_0$ and $m_I$ is found at $\Delta \tilde{t}\approx \,r_0$ in the pure-gauge study, see top plot in Fig.~(4) of Ref.~\cite{Barca:2024fpc}, whereas in this quenched study we find it at $\Delta \tilde{t}\approx 2.3\,r_0$ with $N_1=200$.
This is because the frozen regions are thicker: at fixed $\Delta t$, the effective distance to the boundaries is smaller and boundary fluctuations are larger.
Adapting the crossover estimate of Ref.~\cite{Barca:2024fpc} to frozen regions thicker by $4a$, the transition occurs at
\begin{equation}
    \mathrm{exp}\left(-m^\Gamma (\Delta \tilde{t}-4a)\right) \lesssim \frac{1}{\sqrt{N_1}}\,.
\end{equation}
For instance, the two-level starts to outperform the standard sampling at earlier time separations in the $\Gamma=I$ channel, compared to the $\Gamma=\gamma_5$, because we find that $m^I > m^{\gamma_5}$ at $m_\pi=760~$MeV.
This is a qualitative criterion because correlators/variances are not single-exponential at short-intermediate distances, but it captures that heavier channels enter the two-level regime earlier.
\subsection{Effective masses of singlet observables in quenched QCD}
A preliminary study of the connected and disconnected contributions to the isosinglet two-point functions shows that the connected pieces do not suffer from a severe signal-to-noise problem using standard sampling techniques, whereas most of the statistical noise originates from the disconnected terms.
We therefore compute the connected pieces with standard sampling and add the improved two-level estimators of the disconnected contributions.
For $N_f=3$, the full isosinglet weighted correlators are
\begin{equation}
    \bar{C}_\Gamma(\Delta t) = \bar{C}_\Gamma(\Delta t)_{\rm conn} - 3 \bar{C}_\Gamma(\Delta t)_{\rm disc}\,,
\end{equation}
with disconnected terms as defined in Sec.~\ref{sec:sec2}.
From the weighted averages (Eq.~\eqref{weighted_average}) we form effective masses
\begin{equation}
    m_{\rm eff}(\Delta t) = \mathrm{ln}\left( \frac{\bar{C}(\Delta t)}{\bar{C}(\Delta t+a)} \right)\,.
\end{equation}
\begin{figure}[t]
\centering
\includegraphics[width=\textwidth]{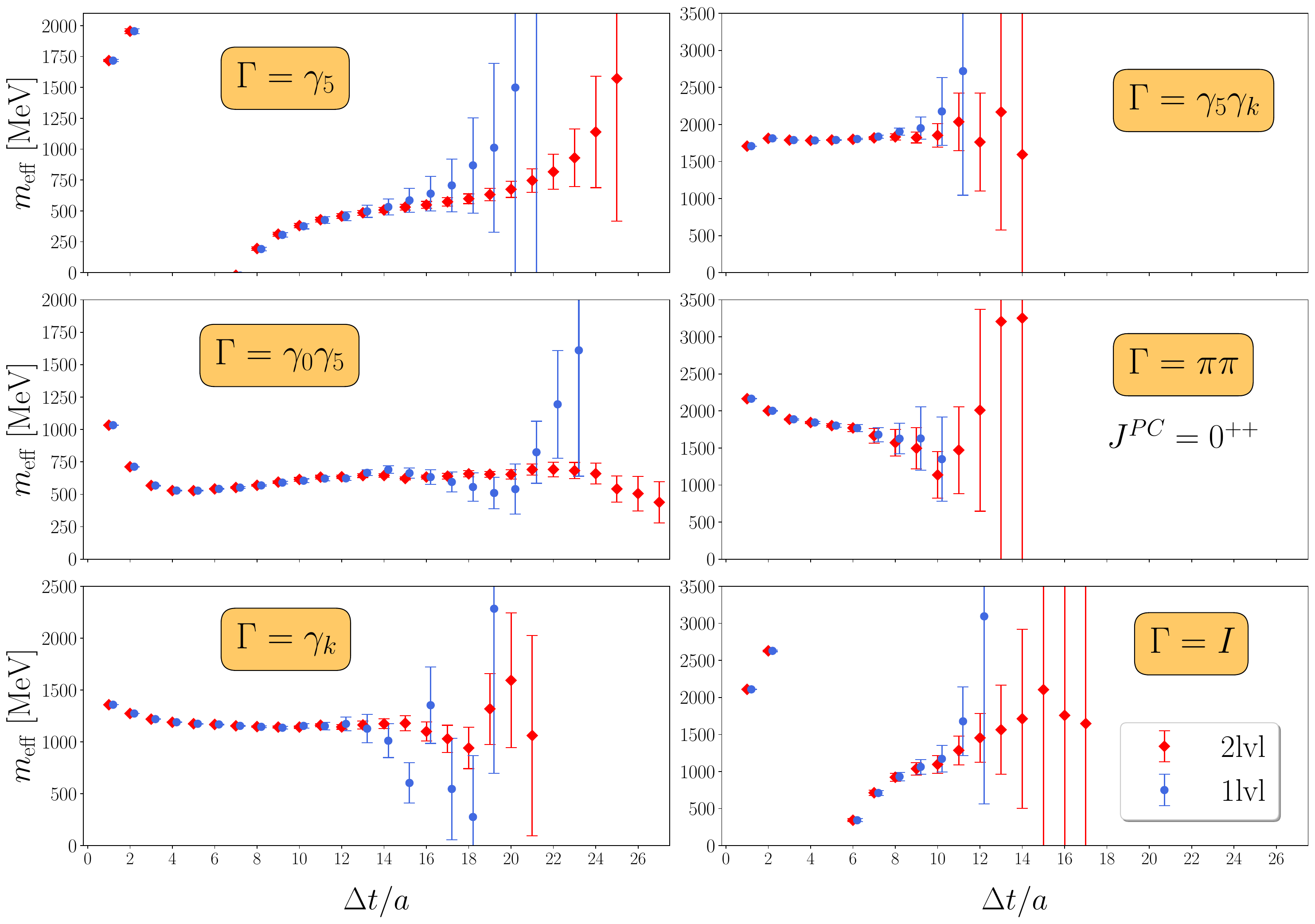}
\caption{Effective masses of singlet two-point functions with $\mathrm{O}_\Gamma=\bar{\psi} \Gamma \psi$ with $\Gamma=\gamma_5, \gamma_4\gamma_5, \gamma_k$, $\gamma_5\gamma_k$, $\gamma_i\gamma_j \epsilon_{ijk}$, and $I$.
}
\label{fig:effective_masses}
\end{figure}
Fig.~\ref{fig:effective_masses} shows $m_{\mathrm{eff}}$ for $\mathrm{O}_\Gamma=\bar{\psi}\Gamma\psi$ with $\Gamma=I,\,\gamma_5,\,\gamma_\mu,\,\gamma_i\gamma_j\epsilon_{ijk},\,\gamma_i\gamma_5$,
comparing one-level ($N_0=101$) and two-level ($N_0=101$, $N_1$ up to $200$) estimators.
The two-level results exhibit substantially improved signal-to-noise for the full isosinglet correlation functions.
We refrain from performing mass fits, since reliable determinations would require a GEVP analysis with a sufficiently large operator basis, and quenched observables are subject to well-known artefacts. 
At large distances in the scalar channel, for example, the dominant state is $\eta'\pi$, whose energy lies below the $f_0$ because the $\eta'$ remains light in the absence of fermion loops; see the partially-quenched discussion in Ref.~\cite{Bardeen:2001jm}. 
The disconnected pseudoscalar channel is likewise affected~\cite{Bernard:1993sv, Sharpe:1996ih}.

\subsection{Cost Budget}
We estimate efficiency in terms of Dirac inversions (dominant cost) and required ensemble size for a given precision.
While the two-level error decreases as $1/N_1$ in amplitude (variance $\sim 1/N_1^2$) until boundary saturation, the one-level error falls only as $1/\sqrt{N_1}$.
Thus, for fixed precision the required number of submeasurements differs.

\begin{figure}%[H]
\centering
%------------------- Top panel -------------------
\begin{subfigure}{\textwidth}
    \centering
    \includegraphics[width=0.92\textwidth]{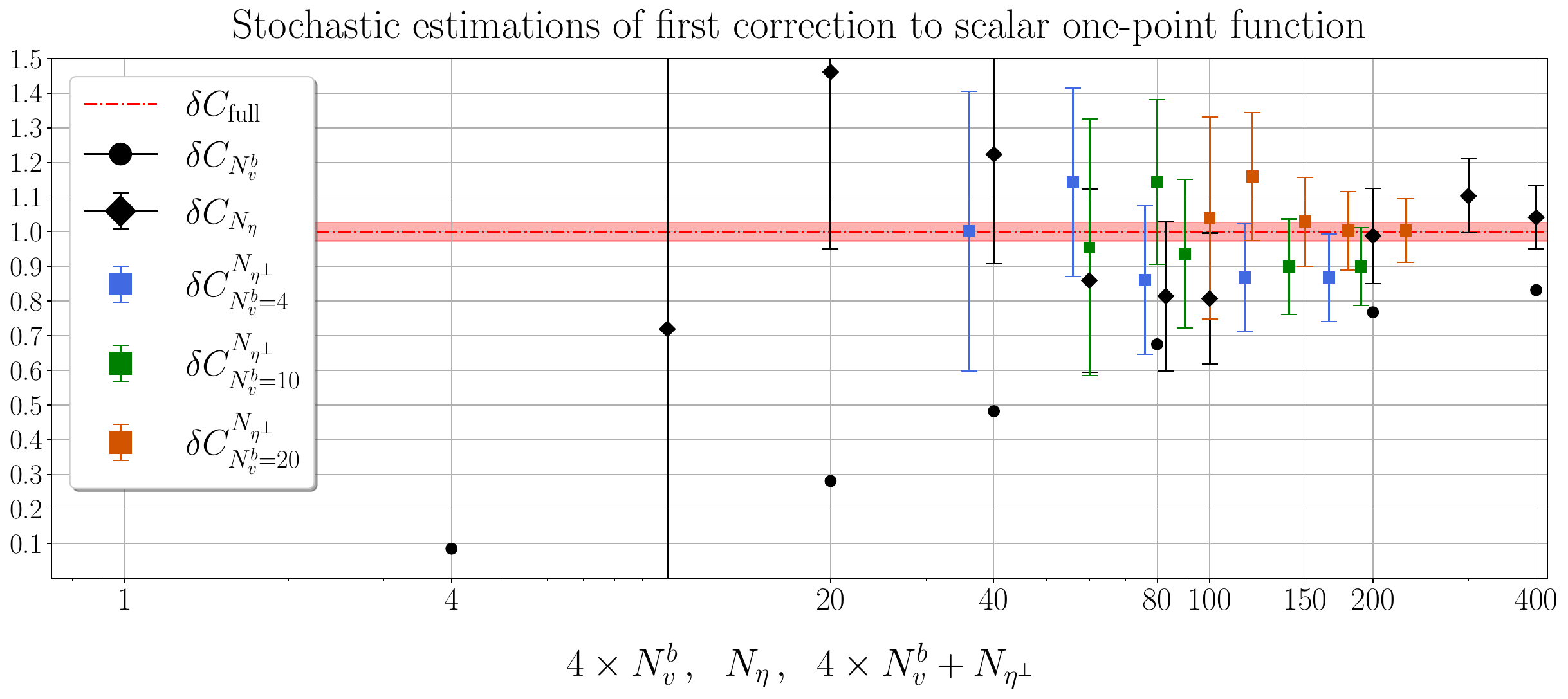}
    \caption{Signal for the normalized estimators of the first correction to the scalar one-point function at $x_0=27a$, $\delta C$, defined as the ratio of the stochastic estimator (Eq.~\eqref{firstcorr_ql_omega1_eta}) to its standard determination (Eqs.~\eqref{firstcorr_ql_omega1}--\eqref{firstcorr_ql_omega0}). The stochastic estimation uses $\tilde{\eta}_k$ taken exclusively as Laplacian eigenvectors (black circles), exclusively as stochastic vectors (black diamonds), or as deflated stochastic vectors (coloured squares).}
    \label{fig:first_corr_1pt}
\end{subfigure}

\vspace{0.4cm}

%------------------- Bottom panel -------------------
\begin{subfigure}{\textwidth}
    \centering
    \includegraphics[width=\textwidth]{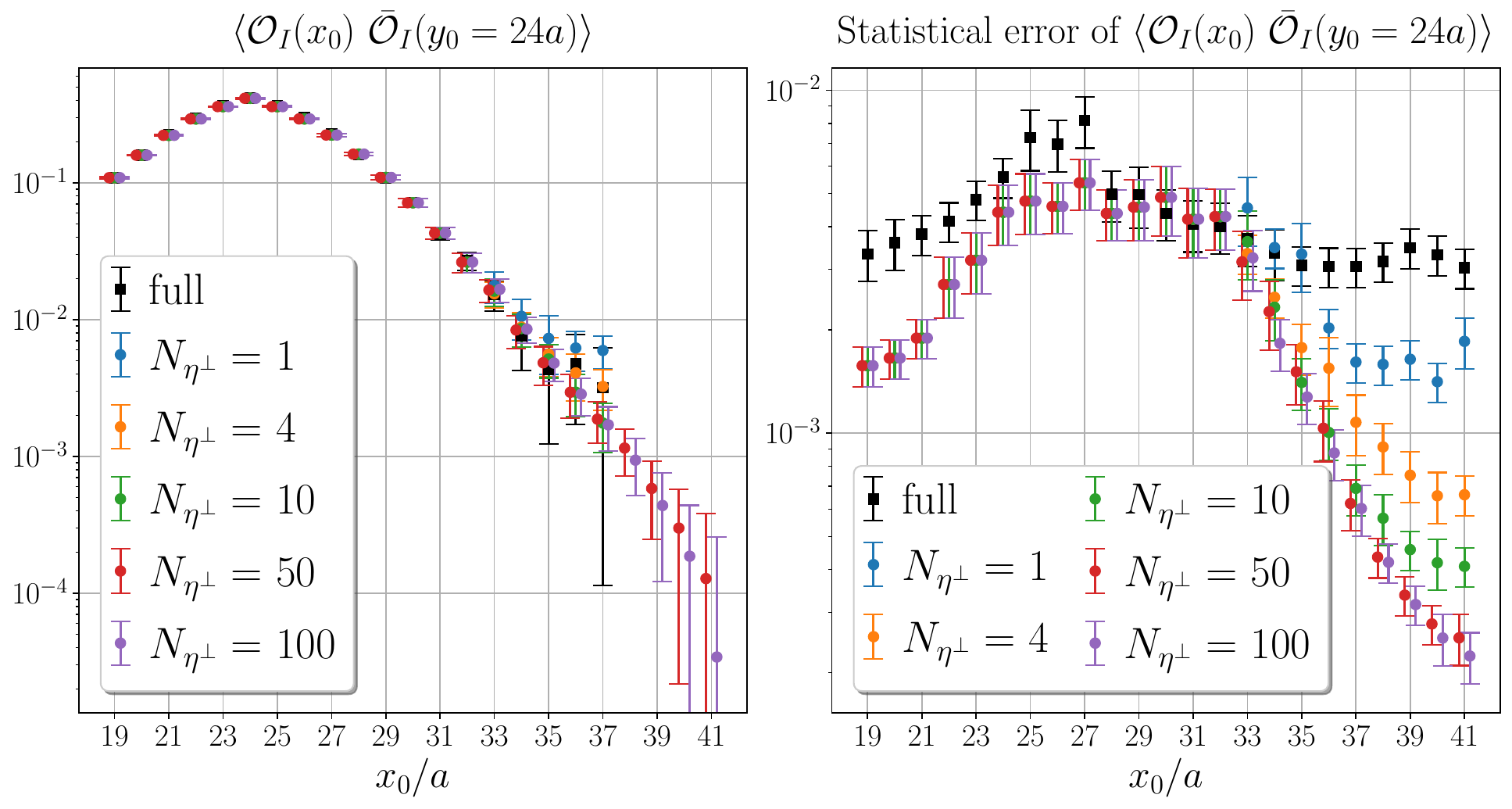}
    \caption{Comparison of the signal for the estimator of the disconnected contribution to the scalar two-point functions between the exact estimator using standard sampling and the deflated stochastic estimator with $N_v^b=10$ using two-level sampling. 
    }
    \label{fig:saturation}
\end{subfigure}

\caption{Analysis of the saturation of stochastic noise for the first correction to one-point function estimator (a) and the two-point function estimator (b).}
\label{fig:combined_firstcorr_2pt}
\end{figure}

In standard sampling, using distillation on $N=N_0N_1^2$ configurations, the cost in inversions scales as $4N_v N_t \times N_0N_1^2$ (factor $4$ for spin indices).
In the two-level approach, we first factorise the propagators and stochastically estimate boundary pieces using $N_{\tilde{\eta}}=4N_v^b + N_{\eta^\perp}$ vectors (e.g.\ $D_{\Lambda_{10}}D^{-1}_{\Omega_0}D_{\Lambda_{12}}\tilde{\eta_k}$ in Eq.~\eqref{psi_k_omega1}), adding $2N_{\tilde{\eta}}$ solves per configuration.
These boundary solves can be reused across sources/sinks positions.
If $N_v^b \leq N_v$, low-mode solutions $D^{-1}_{\Omega_\ell}v_n$ are reused for both the leading term and the first correction, so the $4N_v^b$ solutions are not an extra cost.
To correct the approximation globally (when done by difference with the full solver), one computes the observable with the full propagator: this costs $4N_vN_t$ additional inversions per configuration.

Collecting terms, the additional inversions required by the two-level method are $4N_v N_t \;+\; 2N_{\tilde{\eta}}$
per configuration, i.e.\ a total of $\left(2\cdot 4N_v N_t + 2N_{\tilde{\eta}}\right) N_0N_1$ inversions for leading plus first-correction two-level estimates (counting both domains).
By contrast, the standard method requires $4N_v N_t \times N_0N_1^2$ inversions to reach the same precision at large distances.

For the present parameters $N_v=10$, $N_t=64$, $N_0=101$, $N_1=1000$, $N_{\eta^\perp}=50$, the standard sampling would require
$\sim 2.6\times 10^{11}$ inversions, versus $\sim 2.7\times 10^{8}$ for the two-level method.
Numerically, at $x_0=27a$ in the scalar channel, using $N_v^b=N_v=10$ accounts for roughly 50\% of the first correction (black circles in Fig.~\ref{fig:first_corr_1pt}); the remaining piece is captured with $N_{\eta^\perp}=50$ deflated stochastic vectors.
Hence, the extra inversions to reach the second approximation are modest, while the two-level sampling delivers an almost \(N_1\!\sim\!10^3\) gain in precision at large distances.

In Fig.~\ref{fig:saturation}, we show the signal (left plot) and statistical errors (right plot) of the corrected second-order approximation of the disconnected contribution to the scalar two-point functions with $N_{\eta^\perp}=1,~4,~10,~50,~100$.
We find that $N_{\eta^\perp}=50$ is already sufficient to obtain an accurate stochastic determination of the signal, 
see left plot in Fig.~\ref{fig:saturation}, and to saturate the introduced stochastic noise in the two-level estimation of the first correction, at least up to $x_0=41a$ for $y_0=24a$, $N_0=101$ and $N_1=200$, see right plot in Fig.~\ref{fig:saturation}.

\section{Conclusions}
In this study, we investigate the application of two-level and standard sampling techniques to decrease the error of disconnected fermionic observables in quenched QCD, which hinder the reliable estimation of effective masses of singlet observables. To enable two-level sampling for quark loops, we factorised the quark propagator into contributions depending on two overlapping regions--comprising dynamical and frozen domains, thereby allowing independent sampling of the loop contributions. 
Compared to our previous pure-gauge study~\cite{Barca:2024fpc}, we employed thicker frozen regions at the second level to ensure that correction terms are exponentially suppressed.
In this work, we compute the first correction both with standard and two-level sampling--a key novelty of the present study--while the remaining higher-order corrections are evaluated globally using standard sampling by taking the difference with the correlation functions obtained from full propagators.
We find that the two-level sampling outperforms standard sampling by significantly reducing statistical uncertainties at large distances, thereby enabling the analysis of correlation functions at larger distances. The high-statistics of $N_0=101$ and $N_1=1000$ enables us to study the scaling behaviour of the two-level error reduction, and we find that it is consistent with that observed in the pure gauge case.
In particular, the constant term in the variance decreases as $1/N_1^2$, accompanied by exponential corrections originating from the frozen regions. As the thickness $\Delta$ of the frozen regions is larger than in the pure-gauge setup, where the dynamical domains were separated by only a single time slice, the two-level algorithm begins to outperform at larger separations, reducing the two-level gain in the error.
We also find that efficient scaling of the two-level error reduction at large $N_1$ requires evaluating higher-order terms of the propagator series with a two-level scheme. In practice, for $N_0=101$ and $N_1\approx 200$, the first correction becomes the dominant contribution to the statistical error of the disconnected piece of the scalar two-point functions, see Fig.~\ref{Figure:combined_scalar_2pt}. Computing this correction stochastically with a two-level sampling leads to a clear additional reduction in the variance, and we find that $N_\eta=50-100$ stochastic vectors are already sufficient to saturate the induced stochastic noise for the distances and for the statistics used in this work ($N_0=101,~N_1=200$). 
As $N_1$ increases further, the higher-order terms in the two-point function expansion start to dominate the residual error as they decrease like $\sim 1/\sqrt{N_1}$ at fixed $N_0$. Consequently, these terms must also be computed with a two-level scheme if one aims to maintain the two-level error reduction at larger $N_1$.
It is particularly promising that computing \emph{only} the first approximation of the disconnected two-point function with a two-level scheme already reduces the long-distance error by a factor of~$\approx 200$ with $N_1=200$, as one can see from Fig.~\ref{Figure:combined_scalar_2pt}, 
showing that the leading term in the propagator series captures the dominant source of error.
This study confirms the high efficiency of the two-level compared to the standard sampling and paves the way for applying this sampling technique for dynamical studies.

\begin{acknowledgments}
We thank J.\,A.\,Urrea-Ni\~no and the members of the FOR5269 research unit for sharing the distillation code. 
The authors gratefully acknowledge the Gauss Centre for Supercomputing e.V. (www.gauss-centre.eu) for funding this project by providing computing time on the GCS Supercomputer SuperMUC-NG at Leibniz Supercomputing Centre (www.lrz.de) and the scientific support and HPC resources provided by the Erlangen National High Performance Computing Center (NHR@FAU) of the Friedrich-Alexander-Universität Erlangen-Nürnberg (FAU) under the NHR project k103bf. NHR funding is provided by federal and Bavarian state authorities. NHR@FAU hardware is partially funded by the German Research Foundation (DFG) – 440719683.
The work is supported by the German Research Foundation (DFG) research unit FOR5269 "Future methods for studying confined gluons in QCD".
J.F.~acknowledges financial support by the Eric \& Wendy Schmidt Fund for Strategic Innovation
through the CERN Next Generation Triggers project under grant agreement number SIF-2023-004.
This project has received funding by the European Union. Views and opinions expressed are however those of the authors only and do not necessarily reflect those of the European Union or the European Research Council Executive Agency (ERCEA). Neither the European Union nor the ERCEA can be held responsible for them.
The sole responsibility for the content of this publication lies with the authors.
\end{acknowledgments}

\bibliography{references}

\end{document}